\documentclass{article}

\usepackage{PRIMEarxiv}

\usepackage[utf8]{inputenc} 
\usepackage[T1]{fontenc}    
\usepackage{hyperref}       
\usepackage{url}            
\usepackage{booktabs}       
\usepackage{amsfonts}       
\usepackage{nicefrac}       
\usepackage{microtype}      
\usepackage{lipsum}
\usepackage{fancyhdr}       
\usepackage{graphicx}       
\graphicspath{ {./figures/} }
\usepackage{caption}  
\usepackage{multirow}
\usepackage{listings}
\usepackage{caption}
\usepackage{amsmath}
\usepackage{array}

\lstdefinestyle{R}{
    language        = R,
    basicstyle      = \footnotesize,
    keywordstyle    = \color{blue},
    stringstyle     = \color{cyan},
    commentstyle    = \color{magenta}\ttfamily,
    breaklines=true,
    showstringspaces=false
}

\title{Super-intelligence or Superstition? Exploring Psychological Factors Influencing Belief in AI Predictions about Personal Behavior
}

\author{
  \parbox[t]{0.45\textwidth}{
    \centering
    Eunhae Lee \\
    \textmd{MIT Media Lab \\
    Massachusetts Institute of Technology \\
    Cambridge, MA\\
    \texttt{eunhae@mit.edu}}
    \thanks{\textit{Corresponding author}}
  }
  \And
  \parbox[t]{0.45\textwidth}{
    \centering
    Pat Pataranutaporn \\
    \textmd{MIT Media Lab \\
    Massachusetts Institute of Technology \\
    Cambridge, MA\\
    \texttt{patpat@mit.edu}}
  }
  \And
  \parbox[t]{0.45\textwidth}{
    \centering
    Judith Amores \\
    \textmd{Microsoft Research \\
    Cambridge, MA \\
    \texttt{judithamores@microsoft.com}}
  }
  \And
  \parbox[t]{0.45\textwidth}{
    \centering
    Pattie Maes \\
    \textmd{MIT Media Lab \\
    Cambridge, MA \\
    \texttt{pattie@media.mit.edu}}
  }
}

\begin{document}
\maketitle

\begin{abstract}
Could belief in AI predictions be just another form of superstition? This study investigates psychological factors that influence belief in AI predictions about personal behavior, comparing it to belief in astrology- and personality-based predictions. Through an experiment with 238 participants, we examined how cognitive style, paranormal beliefs, AI attitudes, personality traits, and other factors affect perceived validity, reliability, usefulness, and personalization of predictions from different sources. Our findings reveal that belief in AI predictions is positively correlated with belief in predictions based on astrology and personality psychology. Notably, paranormal beliefs and positive attitudes about AI significantly increased perceived validity, reliability, usefulness, and personalization of AI predictions. Conscientiousness was negatively correlated with belief in predictions across all sources, and interest in the prediction topic increased believability across predictions. Surprisingly, we found no evidence that cognitive style has an impact on belief in fictitious AI-generated predictions. These results highlight the "rational superstition" phenomenon in AI, where belief is driven more by mental heuristics and intuition than critical evaluation. This research advances our understanding of the psychology of human-AI interaction, offering insights into designing and promoting AI systems that foster appropriate trust and skepticism, critical for responsible integration in an increasingly AI-driven world.
\end{abstract}


\section{Introduction}

Artificial Intelligence (AI) is often associated with scientific advancement and frequently viewed in opposition to superstition. Yet, despite its foundations in mathematics and computational logic, the public's engagement with AI may not be as rational or evidence-based as one might expect. In his philosophical paper, Alexander Wilson explores the concept of "rational superstition" as it relates to techno-optimism, challenging traditional ideas of reason and causality. He argues that technological optimism often involves a quasi-superstitious belief in the power of technology to shape the future, which reflects how irrationality can influence our relationship with technological systems  \cite{wilson_techno-optimism_2017}. This behavior, where individuals engage with AI predictions in ways that echo quasi-religious beliefs or divination, has been further observed to be paralleling society's relationship with astrology or religion \cite{uyar_asi_2024, nikolic_ecs-ecrea_2023, lazaro_pouvoir_2018, steyerl_sea_2018, geraci_apocalyptic_2008}, starkly contrasting with the ideal view of AI as a neutral, rational tool designed to aid human decision-making \cite{valtonen_exploring_2022, thompson_ethical_2021}.

Humans rely on a spectrum of cognitive processes, ranging from rapid, intuitive responses (System 1) to slower, reflective reasoning (System 2), both of which can contribute to cognitive biases and mental shortcuts and skew decision-making \cite{kahneman_thinking_2011, evans_two_2009}. The stakes are high when it comes to AI systems---unchecked, this tendency could drive inappropriate reliance on AI in contexts where its limitations and biases could lead to harmful outcomes \cite{mahmud_decoding_2024, johnson2020algorithmic, logg_algorithm_2019, araujo_ai_2020}. The unscrutinized narratives around AI, often evangelized by Silicon Valley executives and popular media \cite{danaher_techno-optimism_2022}, further complicate the picture. From claims of superintelligence to exaggerated portrayals about AI either saving or destroying humanity, these narratives can distort public understanding of what AI can and cannot do \cite{sigal_silicon_2023, brauner2023public, sartori_minding_2023}, fostering irrational trust or fear of AI akin to religious or pseudoscientific beliefs \cite{uyar_asi_2024, geraci_apocalyptic_2008}. Such polarized views hinder the development of an informed, rational relationship with AI, ultimately posing barriers to its effective and ethical integration into society \cite{wilson_techno-optimism_2017}.

Addressing these challenges requires a collective effort from researchers, developers, policymakers, business leaders, and the public. As AI becomes increasingly integrated into everyday life, from recommender systems to personal assistants, fostering critical and informed engagement with AI technologies is essential. Moving beyond simplistic, technology-focused narratives demands a deeper understanding of the psychological and contextual factors shaping people’s perceptions and decisions about AI \cite{sartori_sociotechnical_2022, floridi_ai4people_2018}.

This study aims to shed light on the psychological and contextual factors influencing trust and engagement with predictive systems by examining how individuals evaluate predictions about personal behavior from AI, astrology, and personality psychology in the context of financial decision-making. The context is particularly timely given the growing popularity of astrology among Millennials and Gen Z \cite{das_fixating_2022, page_why_2023, farrar_why_2022}, despite its lack of empirical validity \cite{mcgrew_scientific_1990, nyborg_relationship_2006}. Surveys reveal that a significant portion of Millennials and Gen Z in the U.S. have made financial decisions based on horoscopes \cite{safier_written_2021}, fueling the rise of "financial astrology" as a pseudoscientific trend \cite{pasavento_traders_2015, rogelberg_gen_2024}. 

To explore these dynamics, an experiment was conducted with 238 participants who were presented with fictitious predictions about their personal investing behavior from three sources---AI for its empirical rigor, astrology for its pseudoscientific appeal, and personality psychology as a mix of perceived relatability and scientific grounding. Participants completed fictitious assessments on astrology and personality traits before engaging in a simulated investment game designed to mimic real-world financial decision-making. They then evaluated the perceived validity, reliability, usefulness, and personalization of the predictions. Participants were randomly assigned to receive either positive (N = 119) or negative (N = 119) predictions about their future investment outcomes, allowing us to examine how the valence of predictions influenced their responses.

The study extends prior research by exploring how cognitive style, paranormal beliefs, gullibility, attitudes toward AI, and personality traits influence belief in predictions from these sources. Drawing on insights from work on the effects of user characteristics on the perception and trust of AI-generated advice \cite{wester_exploring_2024, riedl_is_2022}, we consider the interplay between cognitive style and paranormal beliefs, including astrology \cite{pennycook_2012, ballova_mikuskova_effect_2020, bensley_critical_2023, torres_validation_2023, pennycook_everyday_2015}, building on findings about how cognitive style affects belief in (mis)information and conspiracy theories on social media \cite{stecula_social_2021, mosleh_cognitive_2021} and fake news \cite{bronstein_belief_2018}. Demographic factors such as age, gender, education, and prior experience with AI, astrology, and personality psychology, were also analyzed, alongside participants' interest in investing behavior.

Guided by this framework, the study tested the following hypotheses:

\begin{itemize}
    \item H1: Individuals who are more likely to believe in predictions based on astrology and personality will be more likely to believe in AI predictions.
    \item H2: Individuals with a more analytic cognitive style will find AI predictions more credible and reliable than astrology- or personality-based predictions.
    \item H3: Demographic factors and personal attributes (cognitive reflection, trust in AI, paranormal beliefs, gullibility, personality, prior experience, level of interest in investing behavior) will moderate the perceived validity, reliability, usefulness, and personalization of predictions from different sources.
\end{itemize}

Overall, this study examines the psychological and contextual factors that shape trust and engagement with AI systems, offering insights into the "rational superstition" phenomenon and its implications for decision-making in an increasingly AI-driven world. By addressing the interplay between cognitive biases, mental models, and societal narratives, it highlights the need for user-centered design approaches that foster transparency, critical thinking, and accurate mental models. The findings provide actionable strategies for ethical and responsible AI design, public policy, and educational initiatives aimed at promoting AI literacy and informed decision-making. Finally, it underscores the need for interdisciplinary collaboration to design and promote AI systems that align with user needs and societal values, ensuring that AI technologies are integrated responsibly and equitably into everyday life.

\section{Results}

Our results with 238 participants demonstrate that belief in AI predictions is positively correlated with belief in astrology, suggesting that non-rational factors, such as superstitious thinking, significantly influence how individuals engage with AI systems. Participants with higher paranormal beliefs and positive attitudes toward AI were more likely to perceive AI predictions as valid, reliable, useful, and personalized, even when faced with fictitious outputs. Surprisingly, cognitive style did not significantly influence belief in AI, astrology, or personality-based predictions. Additionally, conscientiousness was negatively correlated with belief in predictions across all sources, while interest in investing behavior (prediction topic) increased believability. These findings emphasize the importance of developing better mental models and public education to foster critical and rational AI engagement, rather than reliance on speculative thinking.

\subsection{People who are more likely to believe in astrology and personality-based predictions are more likely to believe in AI predictions.}

To challenge the assumption that people approach AI predictions with more rationality than other speculative forms of prediction, we hypothesized that there would be a correlation between people's perceptions of AI predictions and perceptions of predictions based on astrology and personality (H1). If this were \textit{not} true, we should see \textit{less} correlation between the two---people who find AI predictions more believable would find astrology predictions less believable, and vice versa. 

To test this hypothesis, a multiple linear regression model was applied to the data in wide format, with believability score for AI predictions as the outcome variable and believability scores for astrology- and personality-based predictions as predictor variables, along with control variables (cognitive style, paranormal beliefs, gullibility, AI attitude/trust in AI, big five personality, familiarity with the prediction sources (AI, astrology, personality), interest in topic of prediction, age, gender, education level). Believability scores for each prediction were calculated by averaging across the four subscales (perceived validity, reliability, usefulness, and personalization), which were drawn from existing literature on the different qualities of a statement that contribute to persuasion, technology adoption, and user attitudes (see Section \ref{sec: believability_subscales}.

\begin{figure*}[h]
    \centering
    \begin{minipage}{0.5\textwidth}
        \centering
        \includegraphics[width=0.9\textwidth]{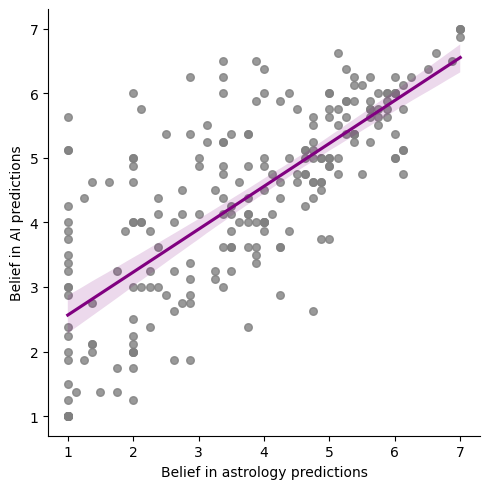} 
    \end{minipage}\hfill
    \begin{minipage}{0.5\textwidth}
        \centering
        \includegraphics[width=0.9\textwidth]{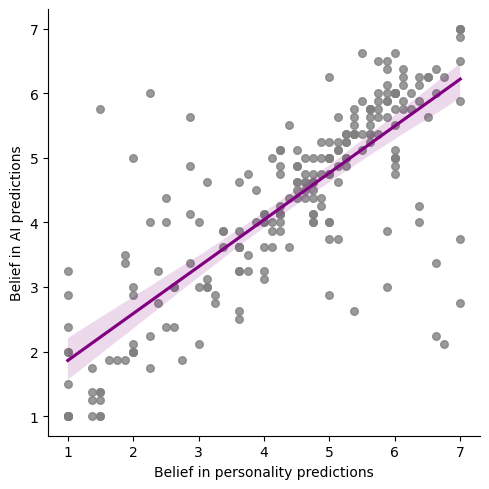} 
    \end{minipage}
    
    \caption{Relationship between belief in AI predictions and \textbf{astrology} predictions (left) and belief in AI predictions and \textbf{personality} predictions (right), measured on a 7-point Likert scale (1 = Strongly disagree, 7 = Strongly agree). Scatterplots display individual scores with linear regression lines indicating positive correlations, and the shaded area represents the 95\% confidence interval for the regression lines.}
    
\label{fig:scatterplots}
\end{figure*}

The results showed a significant positive correlation between belief in AI predictions and belief in astrology- and personality-based predictions (see Figure \ref{fig:scatterplots}), suggesting that people do not engage with AI predictions as rationally as popular narratives might suggest. The multiple linear regression analysis explained a significant proportion of the variance in the AI overall score (\(R^2 = 0.7606\), Adjusted \(R^2 = 0.7337\), \(F(24, 213) = 28.2\), \(p < 0.001\)). Zodiac overall score (Estimate = 0.3119, \(p < 0.001\)) and the personality overall score (Estimate = 0.4585, \(p < 0.001\)) were significant predictors of the AI overall score, supporting the hypothesis that belief in astrology and personality-based predictions is positively associated with belief in AI predictions.

\subsection{People generally find fictitious AI predictions about their personal behavior convincing.}

To examine the influence of cognitive and psychological factors including cognitive style (H2), paranormal beliefs, trust in AI, personality traits, demographic factors, etc. (H3) on belief in AI predictions, a mixed-effects model was fitted to the data in a nested long format, with prediction source (AI, astrology, personality) and believability subscale (perceived validity, reliability, usefulness, and personalization) as categorical predictors, and subscale score as the outcome variable. This approach allowed examination of the main effects and moderating effects of psychological and cognitive factors on belief in AI predictions. 

The model included fixed effects for subscale, prophecy source, prophecy group, cognitive style, paranormal beliefs, gullibility, AI attitude/trust in AI, big five personality, familiarity with sources, interest in topic, age, gender, education level. Interaction terms were included to assess the combined effect of prophecy source and select moderating factors on subscale scores. The model also included random effects to account for individual variability, with the random intercepts capturing the baseline differences among subjects and the random slopes capturing the variability within each participant across different conditions. The inclusion of random intercepts and slopes improved the model fit. For more on our methodology, see Section \ref{sec:methods}.

The results indicate that participants generally found fictitious predictions from AI, astrology, and personality sources to be convincing. The baseline belief score, associated with the AI prophecy source, "Validity" subscale, "Positive" prophecy group, "Female" gender, and "Bachelor's" education level, was 5.10 (p < 0.001) on a 7-point Likert scale. The main effects of prophecy source on the perceived believability of AI predictions revealed that astrology-based predictions were, on average, rated 0.66 points lower in validity compared to AI predictions (95\% CI [-0.93, -0.39], p < 0.001). In contrast, the difference for personality-based predictions was not statistically significant (Estimate = 0.07, 95\% CI [-0.18, 0.32], p = 0.593), suggesting they were perceived similarly to AI predictions.

\begin{figure*}[h]
    \centering
    \includegraphics[width=0.9\textwidth]{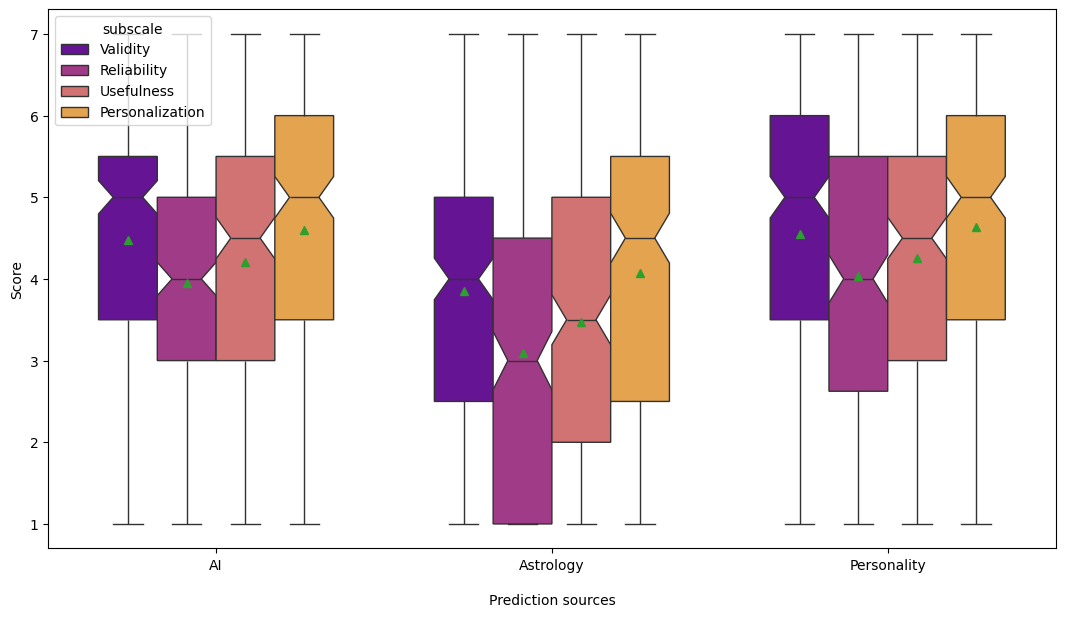} 
    \caption{Boxplot illustrating the distribution of subscale scores (7-point Likert scale: 1 = Strongly disagree, 7 = Strongly agree) across prophecy sources and subscales. The boxes represent the interquartile range, with green triangles indicating the means and notches marking the 95\% confidence intervals for the medians.}
    \label{fig:boxplot}
\end{figure*}

There were significant main effects of subscales on the perception of AI predictions. On average, perceived personalization was rated 0.23 point higher than perceived validity (95\% CI [0.08, 0.38], p = 0.002). Conversely, perceived reliability was rated 0.91 points lower than perceived validity (95\% CI [-1.13, -0.68], p < 0.001), and perceived usefulness was rated 0.63 points lower (95\% CI [-0.83, -0.44], p < 0.001). These relationships between subscales and prediction sources are visualized in Figure \ref{fig:boxplot}. 

The interaction between subscales and prophecy source showed that these main effects were generally consistent across AI, astrology, and personality-based predictions, though some effects were more pronounced for astrology. Specifically, perceived personalization was 0.23 points higher (95\% CI [0.02, 0.44], p = 0.035) and perceived reliability was 0.31 points lower (95\% CI [-0.62, -0.00], p = 0.048) for astrology-based predictions compared to AI predictions. Other interaction terms were not significant, suggesting that the subscale effects were not substantially different across the three prediction sources.

Additionally, compared to the baseline "Positive" prediction group, the "Negative" prediction group saw a significant decrease in perceived validity by 1.19 points  (95\% CI [-1.53, -0.85], p < 0.001). No significant interactions between prophecy group and source were found, indicating that this effect was consistent across AI, astrology, and personality predictions. This suggests that while participants generally found fictitious predictions believable, those who received negative predictions were less likely to find them valid.

\subsection{There is no evidence of correlation between cognitive style and belief in predictions.}

We hypothesized that cognitive style, which has been shown to influence how people process and believe information and make decisions \cite{pennycook_2012, ballova_mikuskova_effect_2020, bensley_critical_2023, torres_validation_2023, pennycook_everyday_2015, bronstein_belief_2018}, would significantly impact how individuals engage with AI predictions compared to astrology- or personality-based predictions (H2). Specifically, we anticipated a positive association between cognitive style and belief in AI predictions, with negative interactions between cognitive style and prediction sources (astrology and personality) relative to AI, which served as the reference category.  

However, the results did not support this hypothesis. The composite cognitive score—derived from the Cognitive Reflection Test (CRT-2) \cite{thomson_investigating_2016} and the Need for Cognition (NFC-6) \cite{lins_de_holanda_coelho_very_2020}—did not significantly increase the perceived validity of AI predictions. While a one-point increase in cognitive score was associated with a 0.13-point increase in perceived validity, this effect was not statistically significant (95\% CI [-0.01, 0.26], p = 0.065). The cognitive score ranged from -4.29 to 2.61 (Mean = 0.0, SD = 1.44).

Interaction terms further supported the lack of evidence of the association between cognitive style and belief in predictions. Higher cognitive scores were significantly associated with decreased perceived reliability (Estimate = -0.11, 95\% CI [-0.20, -0.02], p = 0.021) and usefulness (Estimate = -0.12, 95\% CI [-0.20, -0.04], p = 0.004), compared to the reference subscale of validity. No significant association was found with personalization (Estimate = -0.05, 95\% CI [-0.11, 0.01], p = 0.092). Interactions between cognitive style and prediction source (relative to AI, the reference category) were negative for astrology (Estimate = -0.10, 95\% CI [-0.21, 0.01], p = 0.079) and personality (Estimate = -0.07, 95\% CI [-0.18, 0.03], p = 0.170), though neither interaction was statistically significant. Similarly, three-way interactions between subscales, prediction source, and cognitive score were non-significant.

\begin{figure*}[h]
    \centering
    \includegraphics[width=0.8\textwidth]{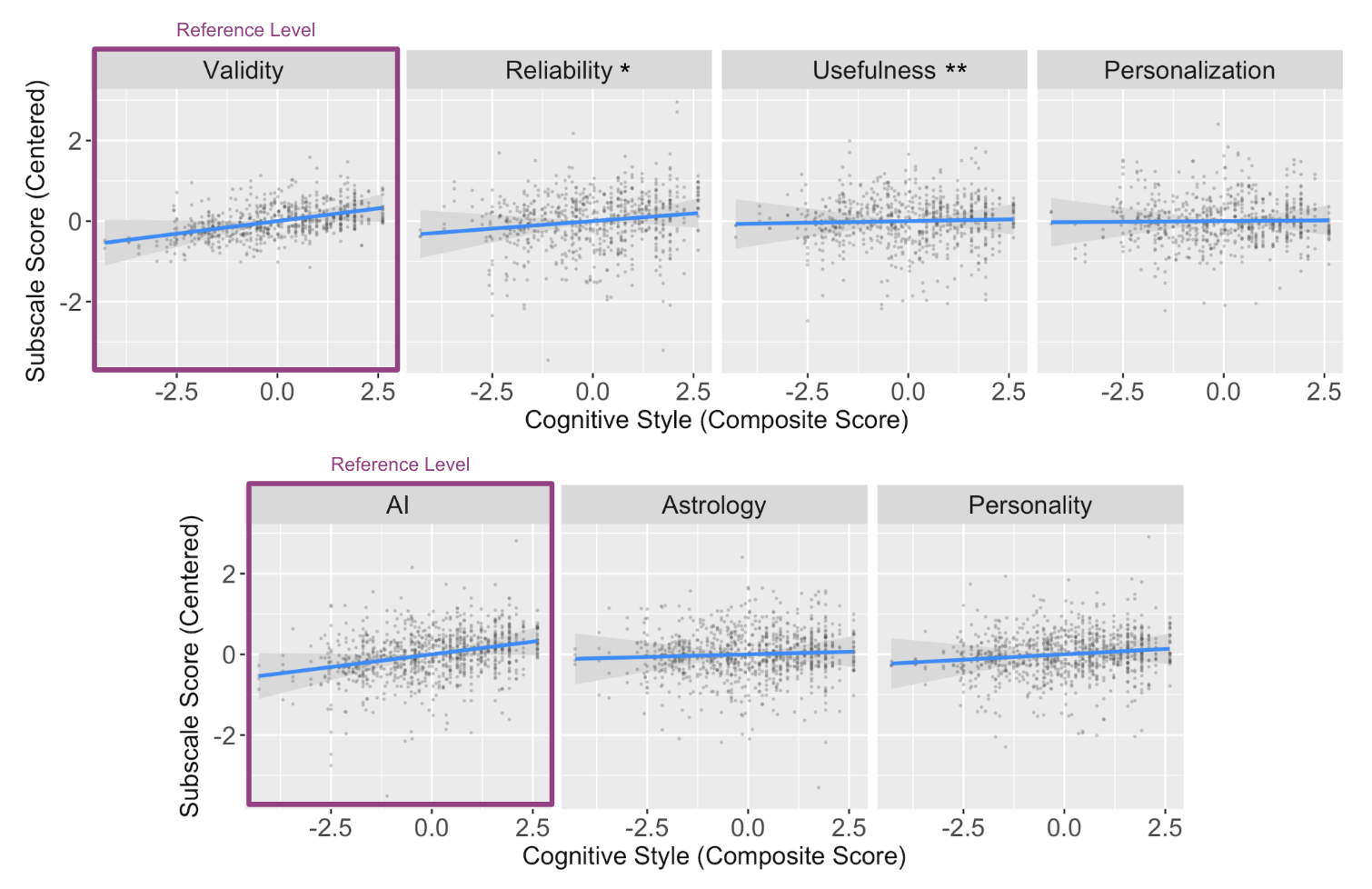} 
    \caption{Contrast plots showing the relationship between \textbf{cognitive style} (Range: -4.29 to 2.61; higher values indicate a more analytic style) and centered subscale scores (7-point Likert scale). The top panel illustrates interactions across subscales (validity, personalization, reliability, usefulness), with "Validity" serving as the reference level. The bottom panel shows interactions across prediction sources (AI, astrology, personality), with "AI" as the reference level. Blue lines represent predicted contrasts with 95\% confidence intervals, and gray points show individual observations. (*p < 0.05, **p < 0.01, ***p < 0.001). Contrast plots are presented for predictors central to the study’s hypotheses. Results for other predictors are summarized in writing within the Results section.}
    \label{fig:contrast_cog}
\end{figure*}

To illustrate these findings, contrast plots in Figure \ref{fig:contrast_cog} present the fixed effects of the model, including main effects and interactions. The top panel visualizes the relationship between cognitive style and subscale scores (validity, personalization, reliability, and usefulness), highlighting overall trends across evaluation dimensions. The bottom panel shows interactions between cognitive style and prediction sources—AI, astrology, and personality—on subscale scores. Contrast plots are presented for predictors central to the study’s hypotheses to visually highlight their effects. Results for other predictors, which also contribute to the findings but are less central to the main hypotheses, are summarized in writing.

\subsection{Higher paranormal beliefs increase perceived validity, reliability, usefulness, and personalization of AI predictions.}

One of the most notable findings was the positive association between paranormal beliefs and belief in AI predictions. Results from the mixed effects model revealed that higher scores on the shortened Revised Paranormal Belief Scale (R-PBS) \cite{tobacyk_revised_2004}, significantly increased the perceived validity of AI predictions. Specifically, each one-point increase in paranormal beliefs led to an average 0.02-point increase in perceived validity (95\% CI [0.01, 0.03], p = 0.001) on a 7-point scale. The paranormal belief scores, centered for analysis, ranged from 15 to 95 (Mean = 45.6, SD = 20.08).

Interactions between paranormal beliefs and subscales showed that this effect was even stronger for perceived reliability (Estimate = 0.01, 95\% CI [0.00, 0.01], p = 0.020) and perceived usefulness (Estimate = 0.01, 95\% CI [0.01, 0.02], p < 0.001). However, there was no significant effect on perceived personalization (Estimate = 0.00, 95\% CI [-0.00, 0.01], p = 0.427), indicating that the impact of paranormal beliefs on perceived personalization did not differ significantly from perceived validity.

\begin{figure*}[h]
    \centering
    \includegraphics[width=0.80\textwidth]{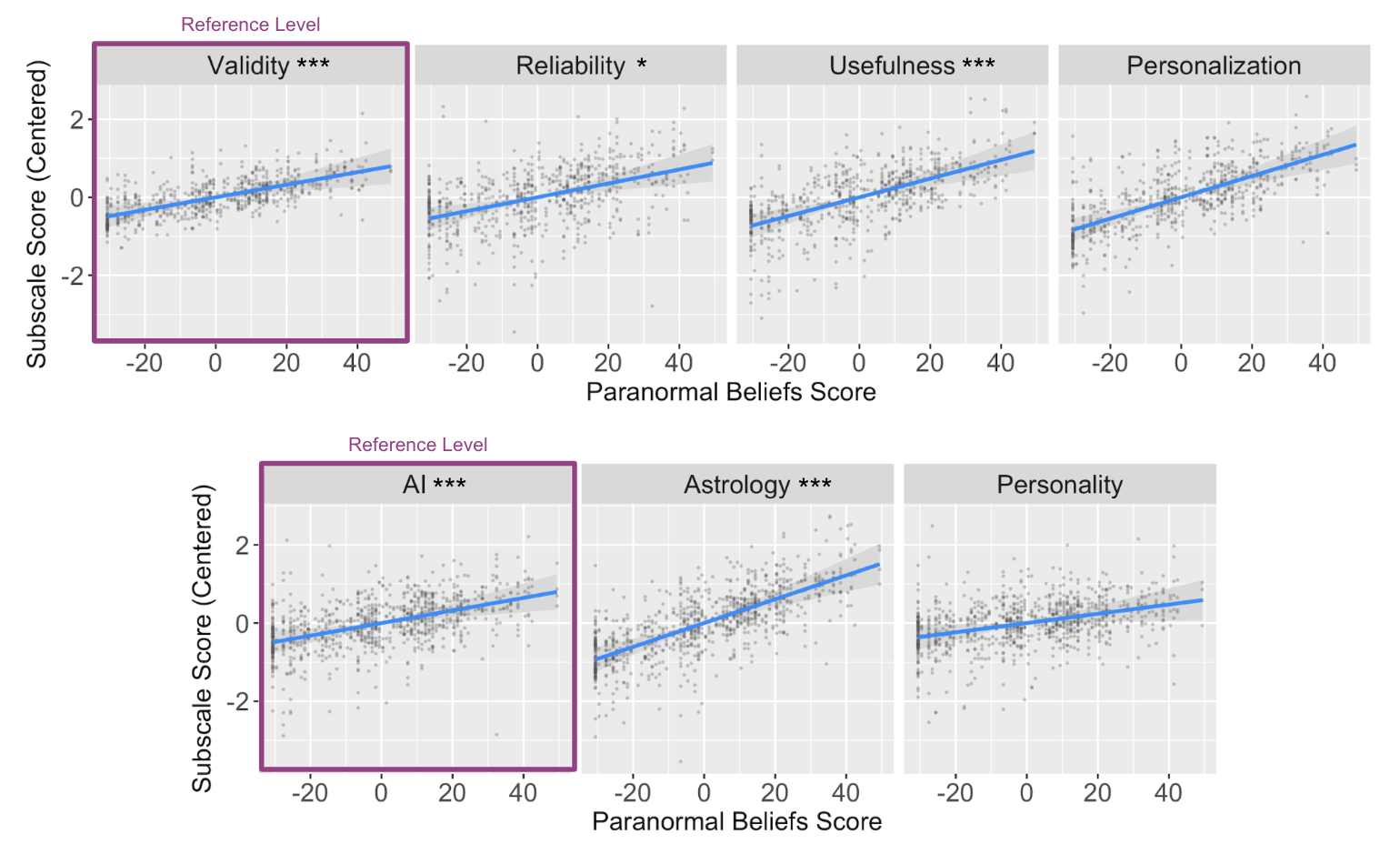} 
    \caption{Contrast plots showing the relationship between \textbf{paranormal beliefs} (higher values indicate stronger paranormal beliefs) and centered subscale scores (7-point Likert scale). Top: interactions across subscales (validity [reference level], personalization, reliability, usefulness). Bottom: interactions across prediction sources (AI [reference level], astrology, personality). Blue lines represent predicted contrasts with 95\% confidence intervals, and gray points show individual observations. The plot ratio was adjusted to align with other contrast plots for visual comparability. (*p < 0.05, **p < 0.01, ***p < 0.001)}
    \label{fig:contrast_para}
\end{figure*}

Interestingly, paranormal beliefs were an even stronger predictor of belief in astrology-based predictions, which aligns with the nature of the scale, as it includes questions about astrology (see Section \ref{sec:scales}). Each point increase in paranormal beliefs led to an additional 0.01-point increase in perceived validity for astrology predictions (95\% CI [0.01, 0.02], p = 0.001) compared to AI. In contrast, differences in personality-based predictions were not significant when compared to AI predictions (Estimate = -0.00, 95\% CI [-0.01, 0.00], p = 0.274). No significant interactions were observed between the subscales, prophecy source, and paranormal score. The contrast plots in Figure \ref{fig:contrast_para} visualize these results.

\subsection{Positive attitudes toward AI increase belief in AI predictions, especially perceived reliability.}

Individuals with more positive attitudes towards AI found AI-based predictions more believable. Each one-point increase in the AI Attitude Scale (AIAS) \cite{grassini_development_2023} was associated with a 0.04-point increase in the perceived validity of AI predictions (95\% CI [0.01, 0.06], p = 0.001) on a 7-point scale. The AIAS score ranged from 4 to 40 (Mean = 25.79, SD = 8.68).

This effect was particularly pronounced for perceived reliability, which increased by an additional 0.03 points (95\% CI [0.02, 0.05], p < 0.001), while the interactions were not significant for perceived usefulness (Estimate = 0.01, 95\% CI [-0.00, 0.03], p = 0.071) or perceived personalization (Estimate = -0.00, 95\% CI [-0.01, 0.01], p = 0.941). This suggests that positive attitudes towards AI are most strongly associated with higher perceived reliability, aligning with previous studies on the relationship between trust in AI and reliance on its predictions \cite{kahr_understanding_2024, klingbeil_trust_2024, chiou_trusting_2023, kuper_psychological_2023, lee_trust_2004}.

\begin{figure*}[h]
    \centering
    \includegraphics[width=0.80\textwidth]{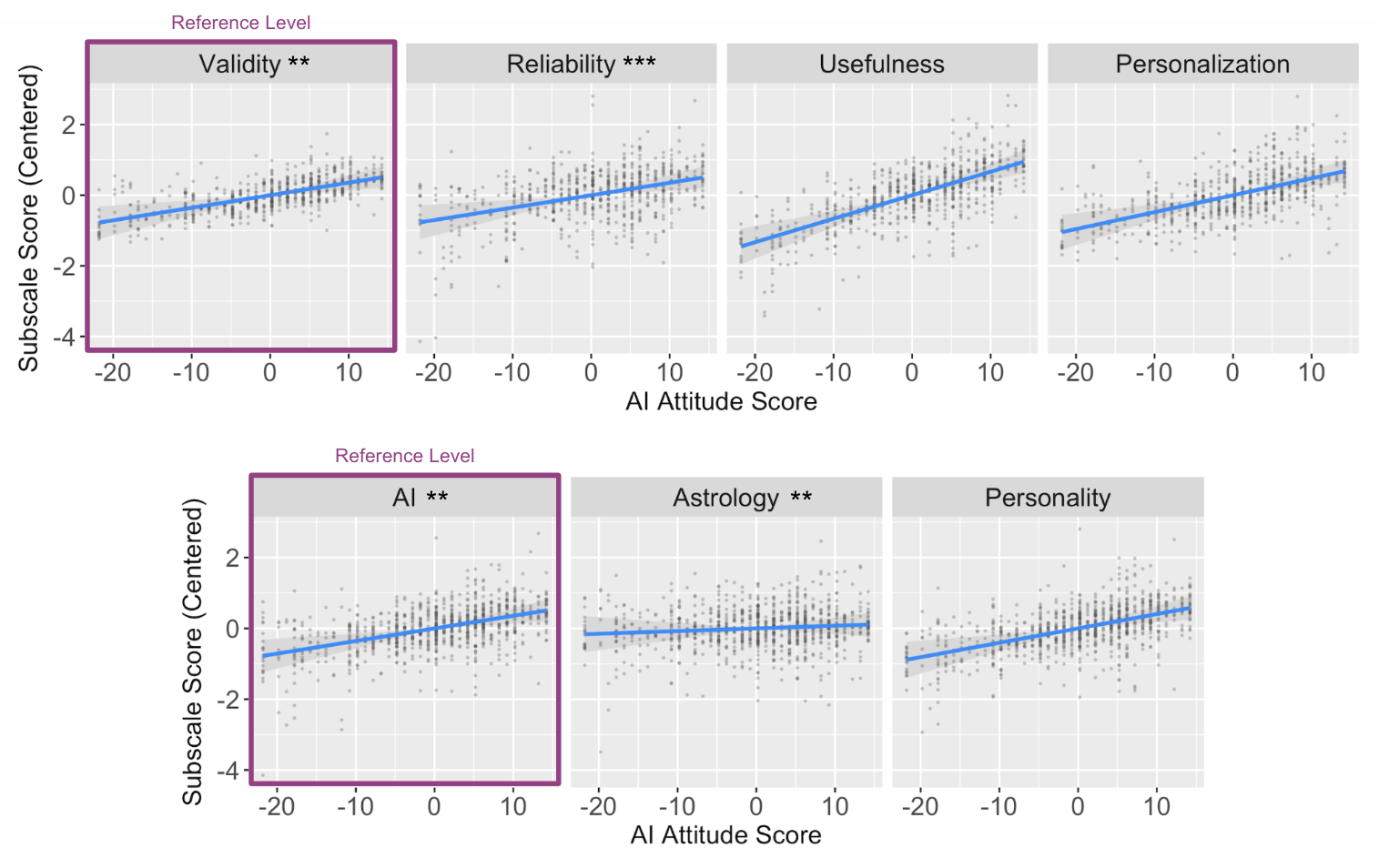} 
    \caption{Contrast plots showing the relationship between \textbf{attitude toward AI} (higher values indicate more positive attitudes) and centered subscale scores (7-point Likert scale). Top: interactions across subscales (validity [reference level], personalization, reliability, usefulness). Bottom: interactions across prediction sources (AI [reference level], astrology, personality). Blue lines represent predicted contrasts with 95\% confidence intervals, and gray points show individual observations. The plot ratio was adjusted to align with other contrast plots for visual comparability.  (*p < 0.05, **p < 0.01, ***p < 0.001)}
    \label{fig:contrast_aias}
\end{figure*}

The interaction between AI attitudes and prediction source revealed that the positive impact of positive AI attitudes on perceived validity was largely reversed for astrology-based predictions, which saw a 0.24-point decrease (95\% CI [-0.41, -0.08], p = 0.004). For personality-based predictions, the effect did not differ significantly from AI predictions (Estimate = 0.04, 95\% CI [-0.11, 0.19], p = 0.590). There were no statistically significant interaction effects between subscales, prophecy source, and AI attitude score. These findings are visualized in the contrast plots in Figure \ref{fig:contrast_aias}.

\subsection{People with high conscientiousness are less likely to believe in predictions about personal behavior.}

Out of the big five personality traits (Extraversion, Openness, Agreeableness, Conscientiousness, Emotional stability), conscientiousness was negatively associated with perceived validity of predictions across all sources. With each point increase in the conscientiousness score (Mean = 5.29, SD = 1.32, Range = [1.0, 7.0], perceived validity was estimated to decrease by an average of 0.15 points (95\% CI [-0.30, -0.01], p = 0.032) on a 7-point scale. There were no significant interaction effects with subscales observed, suggesting the main effect was consistent across the four subscales.

While the other domains of the five-factor model were not found to have significant influence, there were some variation in interaction terms. Extraversion was associated with a 0.06 point increase in perceived usefulness (95\% CI [0.02, 0.10], p = 0.004), while Openness was associated with a 0.06 decrease in perceived personalization (95\% CI [-0.10, -0.01], p = 0.012).

\subsection{Level of interest in the prediction topic increases perceived validity, reliability, usefulness, and personalization.}

While often overlooked in the context of studying trust in AI predictions, individuals' interest in the topic of prediction (in our case, personal investing behavior) was an influential factor in perceived validity, personalization, reliabiliity, and usefulness of predictions. With each point increase in the level of interest in the topic, perceived validity significantly increased on average by 0.27 points (95\% CI [0.11, 0.43], p = 0.001) on a 7-point scale. The interest in behavior scale ranged from 1 to 5 (Mean = 3.18, SD = 1.09).

The positive association was observed across other subscales. Compared to perceived validity, perceived personalization increased slightly less, reduced by 0.06 points (95\% CI [-0.11, -0.01], p = 0.018), while interaction was not statistically significant for perceived reliabilty (Estimate = -0.07, 95\% CI [-0.14, 0.01], p = 0.068) and perceived usefulness (Estimate = 0.02, 95\% CI [-0.04, 0.09], p = 0.475), suggesting that the effects were comparable to perceived validity. The results suggest that when all else is held constant, the more people are interested in the topic, the more likely they will perceive fictitious predictions to be valid, reliable, useful, and personalized.

\subsection{There is no evidence that the level of familiarity with the prediction sources influences belief in predictions.}

On the other hand, familiarity with the prediction sources (AI, astrology, personality psychology), which measures the level of self-reported familiarity/prior knowledge in each source, did not have a significant effect on perceived validity (Estimate = -0.02, 95\% CI [-0.11, 0.07], p = 0.650) across all sources. The familiarity scale ranged from 1 to 5 (Mean = 2.96, SD = 1.05).

The interaction between familiarity and subscales shows that perceived personalization increases slightly but significantly by 0.05 points (95\% CI [0.00, 0.11], p = 0.035), while differences were not statistically significant for perceived reliability (Estimate = 0.02, 95\% CI [-0.06, 0.10], p = 0.685) and perceived usefulness (Estimate = 0.01, 95\% CI [-0.06, 0.08], p = 0.781).

Unlike what the literature suggests about the familiarity effect, a cognitive phenomenon where people tend to prefer things they are familiar with \cite{hansen2009liking}, our results showed inconclusive evidence that familiarity in the prediction sources (AI, astrology, and personality) neither increased nor decreased belief in the respective predictions.

\subsection{Other results}

Gullibility was not found to be a significant predictor of perceived validity of AI predictions (Estimate = -0.02, 95\% CI [-0.04, 0.01], p = 0.128). While the interaction  between gullibility and Personalization subscale was significant (Estimate = 0.01, 95\% CI [0.00, 0.02], p = 0.021), the results were inconclusive when compared to the main effects. Interactions between gullibility and prophecy source were not significant (Astrology: Estimate = -0.00, 95\% CI [-0.02, 0.02], p = 0.977, Personality: Estimate = -0.02, 95\% CI [-0.04, 0.00], p = 0.096), suggesting the effects were similarly insignificant across different prophecy sources. Gullibility scale ranged from 6 to 39 (Mean = 14.68, SD = 7.69). 

We found that older age was associated with a decrease in belief in predictions across all sources. One year increase in age (Mean = 41.04, SD = 12.38, Range = [19, 75]) was associated with a decrease of perceived validity of AI predictions by 0.02 points (95\% CI [-0.03, -0.00], p = 0.014). Interactions between age and subscales were not significant, nor were interactions between age and prophecy source. The non-significant interaction terms suggested that the effect was consistent across subscale and prophecy source.

For gender, we did not find significant main effects relative to the Female reference level. While male participants were associated with lower perceived validity scores for AI predictions than female participants, it was not statistically significant (Estimate = -0.31, 95\% CI [-0.69, 0.06], p = 0.103). However, they tend to perceive them as more reliable than female participants (Estimate = 0.41, 95\% CI [0.14, 0.68], p = 0.003). Interactions with other subscales were not significant, suggesting they were similar to the perceived validity baseline. Interactions between gender and prophecy source were not significant.

There were no significant main effects for level of education compared to the Bachelor reference level. Interactions between education level and subscale showed some significant effects between High school and Personalization (Estimate = -0.23, 95\% CI [-0.40, -0.06], p = 0.009), and Less than high school and Reliability (Estimate = -0.89, 95\% CI [-1.56, -0.22], p = 0.009), but these results did not lead to conclusive results.

\section{Discussion}

Belief is subjective and context-dependent; it is also socially constructed through external stimuli such as media narratives, contextual factors, and internal disposition \cite{berger1966social, oliver2019media}. When looking at belief, trust, and reliance in AI predictions, it is important to consider not just the features of the AI system, but also the characteristics of the users and the broader context. The results from our study show that people who are more likely to believe in astrology- and personality-based predictions were more likely to believe in AI predictions. Moreover, we found that believability of AI predictions on personal behavior can be influenced by various psychological and contextual factors that also drive belief in astrology- or personality-based predictions. 

We conceptualized believability using four subscales---perceived validity, reliability, usefulness, and personalization---each grounded in theoretical and empirical research on persuasion, technology acceptance, and user attitudes (see Section \ref{sec: believability_subscales} for details). This conceptualization provides a more comprehensive framework for understanding how users evaluate the credibility of predictions, offering insight into the psychological and contextual factors that influence decision-making.

\textbf{Cognitive style.} Our findings challenge the assumption that an analytic cognitive style leads to greater skepticism toward predictions based on astrology or personality assessments and a preference for AI-generated predictions. Contrary to prior research \cite{bensley_critical_2023, pennycook_everyday_2015, pennycook_2012} and our hypothesis (H2), cognitive style was not a significant predictor of belief in any type of prediction, including those from AI. Moreover, we observed no inverse relationship between belief in AI and astrology predictions, suggesting that analytic thinkers do not inherently dismiss pseudoscientific forecasts or display heightened trust in AI predictions. This highlights a gap between cognitive style and rational skepticism, pointing to a potential missing mechanism influencing belief in AI predictions. Interestingly, the significant interaction between cognitive scores and perceptions of reliability and usefulness nearly reversed the direction of the nonsignificant main effect (reference level validity), indicating that the impact of cognitive style may depend on context-specific factors, such as how predictions are evaluated for their reliability and usefulness, rather than reflecting a uniform influence across all prediction types.

\textbf{Paranormal beliefs.} Paranormal beliefs emerged as a strong predictor of belief in predictions across all sources, including AI. Although this association may initially seem unexpected, it can be understood within the context of popular narratives about AI. Despite AI being portrayed as rational and objective, which might lead to the assumption that scientifically inclined individuals would find AI predictions more credible, our findings indicate otherwise. Instead, belief in AI predictions was more closely linked to paranormal beliefs than to cognitive style. This aligns with the concept of "rational superstition" in AI, as discussed in the introduction. Furthermore, the influence of paranormal beliefs was significant across all subscales of AI predictions but was especially pronounced for perceived usefulness and reliability. Social science literature suggests that the growing interest in astrology reflects increasing uncertainties and anxieties stemming from the breakdown of traditional communities \cite{bauer_belief_1997}. People may turn to pseudoscientific approaches like astrology for guidance amid these uncertainties. Our findings indicate that individuals might view AI predictions similarly to astrology-based predictions in the context of personal behavior forecasting, offering a potential explanation for the observed results.

\textbf{Attitude toward AI.}
Our results showed that people with positive attitudes towards AI found AI predictions more valid, reliable, useful, and personalized. Among the subscales, perceived reliability was found to be most closely related to attitude towards AI. This supports findings from prior literature that positive attitudes toward AI lead to higher reliance in AI predictions \cite{solberg_conceptual_2022, lee_trust_2004}.

\textbf{Personality.}
Our findings using the Big Five personality traits \cite{mccrae_validation_1987} show that conscientiousness had a negative influence on perception of validity, personalization, reliability, and usefulness, while other traits did not have significant effects. This contrasts with prior literature on personality's influence on trust in AI systems that show that agreeableness, openness, and extraversion generally have a positive influence on trust, while high neuroticism tends to negatively affect trust \cite{riedl_is_2022}. Results on conscientiousness have been inconclusive, with some studies reporting a positive association \cite{bawack_exploring_2021, chien_relation_2016, rossi_impact_2018}, while others indicate a negative relationship \cite{aliasghari_effect_2021, oksanen_trust_2020}.

Conscientiousness is defined as “socially prescribed impulse control that facilitates task- and goal-directed behavior, such as thinking before acting, delaying gratification, following norms and rules and planning, organizing, and prioritizing tasks” \cite[p.~120]{john_paradigm_2008}. 

While recent studies suggest a positive relationship between conscientiousness and cognitive ability \cite{corbeanu_conscientiousness_2023, meyer_conscientiousness_2024}, our results did not Contrary to our expectations that cognitive style and conscientiousness may mirror each other in direction, it was not the case---cognitive style did not have significant effect on believability in our study, while conscientiousness had a significant negative influence on perception of validity, personalization, reliability, and usefulness.

\textbf{Interest in the topic of prediction.} 
Our findings show that people's interest in the topic of prediction is a strong predictor when it comes to their belief in predictions, regardless of the source. A potential explanation is that the more interest an individual has about the topic, the more exposure they may have to information about it, and the more likely they may be susceptible to cognitive biases that could influence their belief. Some biases that could explain this tendency are confirmation bias and belief bias. Confirmation bias is when individuals seek out and remember information that confirm their existing beliefs \cite{pohl_cognitive_2012}. Belief bias occurs when individuals judge the strength of an argument based on the believability of its conclusion rather than the logical structure of the argument itself \cite{evans_conflict_1983, klauer_belief_2000}. 

This has implications in the context of AI-assisted decision-making in different fields; e.g. in marketing, companies may take advantage of people's interest in a certain topic and expose them to more targeted information (e.g. advertisements) that may create an impression of hyper-personalization and lead to more positive attitudes toward the brand \cite{chandra2022personalization}. Moreover, in the medical field, an individual may be more prone to believing AI-based predictions about their health if it confirms their prior beliefs. 

\textbf{Familiarity/Prior knowledge.}
Familiarity with the prediction source was not found to be a significant predictor of belief in predictions. Earlier, we discussed how people's positive AI attitudes and interest in the topic of the prediction were positively correlated with belief in the predictions, which may be explained by its likelihood to fuel certain cognitive biases such as belief bias. On the other hand, an individual's level of familiarity in AI, astrology, and personality psychology may not be necessarily indicative of their attitude towards the field. For instance, someone who is not acquainted with the details of the latest AI developments may have either a utopian or dystopian view of AI's impact on society, as mentioned in the introduction. Similarly, just because an individual is familiar with astrology does not inherently mean that they are more likely to believe or adopt the predictions, as it may be influenced by other personal and contextual factors. This suggests two things: 1) simply knowing more about the prediction source does not predict whether one finds a prediction more or less believable, and 2) it does not make one better at calibrating one's trust, providing implications for the design of trustworthy AI systems.

\textbf{Gullibility.}
While prior literature suggested a positive relationship between gullibility and paranormal beliefs \cite{bensley_critical_2023, torres_validation_2023}, our findings showed inconclusive results on whether self-reported gullibility had a positive or negative effect on the belief of fictitious predictions based on AI, astrology, and personality. There may be a few potential explanations for this. One is that the order of the study potentially biased participants' answers. The survey was completed after participants had seen and evaluated the predictions, at which point their perception of the believability of the predictions may shadow their self-perception of gullibility---in other words, they did not want to contradict their decisions so quickly by admitting that they are more prone to being fooled. Another possible explanation is the limitation of a self-report nature of the gullibility scale. Some people may be hesitant to admit that they are gullible for different reasons. As such, future studies can further explore the relationship between gullibility and belief in AI predictions.

\textbf{Demographic factors.}
Our results found that older people are more skeptical of predictions about personal behavior, across AI, astrology, and personality sources. This is aligned with prior literature that seems to suggest that older age is associated with lower trust in AI \cite{chu_age-related_2023}, due to many reasons including bias, lack of learning avenues, and concerns about privacy \cite{shandilya_understanding_2024}. For gender, while the main effects were insignificant, interactions revealed that male participants were more likely to perceive AI predictions as reliable than female participants. This is supported by a prior study that found that being male with higher education and with a Western background were predictors of trust in AI among the general population \cite{yakar_people_2022}. Moreover, while prior work suggested a positive association between level of education and belief in AI \cite{chu_age-related_2023}, we did not find significant main effects for level of education, and the interaction terms did not lead to conclusive results.

\section{General Discussion}

Our results highlight the irrational ways humans perceive and trust AI predictions, showing that belief in AI predictions may be more influenced by paranormal beliefs and positive attitudes toward AI than by cognitive style or familiarity with the technology. This suggests that highly accurate performance is not necessary for users to trust AI systems, pointing to a disconnect between perception and actual capability. This section discusses practical implications of the results.

\subsubsection*{1) Incomplete mental models of AI can skew perception of validity, reliability, usefulness, and personalization.}

Our findings demonstrate that people’s belief in AI predictions is often shaped by their mental models rather than the actual performance of the AI system. Even with entirely fictitious AI predictions, participants rated them as valid, reliable, useful, and personalized. Building on previous research in the placebo effect of AI, this reflects the significant influence of framing and presentation: users may trust AI based on how it is described rather than its true capabilities \cite{kloft2024ai, kosch_placebo_2022, villa_placebo_2023, villa_evaluating_2024, pataranutaporn2023influencing, nourani_anchoring_2021}. Popular narratives that depict AI as omniscient or highly rational contribute to unrealistic mental models, undermine critical thinking and agency when interacting with the predictions \cite{uyar_asi_2024, crawford_atlas_2021}.

This finding has practical implications for both AI design and public policy. AI systems should be designed to help users develop accurate mental models by using clear, objective language and offering transparency about the system’s limitations. Overhyped marketing or media portrayal can exacerbate misunderstandings about the actual capabilities of AI systems, leading to misplaced trust. Thus, it is important to support AI literacy initiatives that continually educate the public on realistic AI capabilities and limitations, as well as increase efforts to reduce misleading claims. As technology evolves, users’ mental models must also adapt, incorporating a healthy level of skepticism to foster informed, rational interactions \cite{hoff_trust_2015}.

\subsubsection*{2) Cognitive biases can reinforce flawed mental models of AI systems.}

Cognitive biases provide valuable insight into understanding irrational beliefs about AI. Belief bias is one such cognitive bias, where individuals evaluate information based on prior beliefs rather than logical validity \cite{evans_conflict_1983, klauer_belief_2000}. While primarily driven by fast, intuitive System 1 thinking, particularly under time pressure \cite{evans_rapid_2005}, belief bias can also involve slower, more reflective System 2 processes, which may rationalize or reinforce mental models aligned with existing beliefs \cite{de_neys_conflict_2017, evans_two_2009}. Known as the selective processing model, this helps explain why people may fail to engage logical thinking when a statement aligns with their beliefs, especially concerning self-related beliefs \cite{evans2000thinking}.

The Barnum effect, where individuals believe vague, general statements to be highly accurate for them personally, also contributes to this phenomenon \cite{forer_fallacy_1949}. Commonly observed in astrology and personality assessments \cite{glick_fault_1989, johnson_barnum_1985}, the Barnum effect may explain why users perceive generic AI-generated predictions as personalized and valid. These cognitive biases, including belief bias and the Barnum effect, shape and reinforce flawed mental models of AI, leading to overreliance on its outputs \cite{dale_heuristics_2015, logg_algorithm_2019, mahmud_decoding_2024, klingbeil_trust_2024, steyvers_three_2023}. In turn, these incorrect mental models can reinforce cognitive biases, creating a feedback loop that distorts rational engagement with AI.

These cognitive biases highlight why transparency in AI is not enough. While interpretability and explainability efforts aim to make AI systems clearer, explanations that simply confirm users’ existing biases can amplify over-trust rather than promote understanding \cite{danry2022deceptive, gonzalez_interaction_2021}. Effective explanations could go beyond clarifying AI processes to encourage thoughtful, critical engagement with AI outputs \cite{danry2023_ask_me}.

\subsubsection*{3) Promoting critical engagement with AI.}

As our reliance on AI-assisted decision-making grows, our capacity for independent critical thinking may be compromised \cite{chiriatti_case_2024}. To counteract this, AI systems must be designed with an awareness of human mental models and cognitive biases to support the development of an appropriate level of trust and foster critical engagement. 

Recent research in human-AI collaboration propose promising directions. For instance, interactive feedback mechanisms that offer counterfactual scenarios \cite{cheng2024interactive} or pose follow-up questions could prompt users to reconsider their assumptions. Deliberation nudges, such as a "cool-off" period for high-stakes decisions, could also slow down quick judgments and encourage more thoughtful engagement \cite{li_decoding_2024}. Additionally, "seamful" design approaches that strategically exposes mismatches and mistakes of AI "to support user agency, re-configuration, and appropriation" \cite{ehsan_seamful_2024}. Moreover, presenting uncertainty or confidence levels with AI predictions can encourage users to critically evaluate the outputs rather than blindly accept them \cite{prabhudesai2023}. Robust evaluation frameworks and benchmarks should be developed to validate the effectiveness of these approaches in real world systems.

From a policy standpoint, advancing AI literacy is essential to equip the public with the tools to critically engage with AI systems. Initiatives should include clear, accessible guidelines that foster public awareness of the implications of AI-augmented cognition and encourage informed discourse.

The notion that humans are inherently irrational while machines are rational is a flawed perspective. Both humans and AI can exhibit rational and irrational behaviors, and human-machine interactions are shaped by a complex interplay of these tendencies. The goal of AI design should be to support informed and critical decision-making, helping users navigate these complexities in ways that enhance well-being and minimize risks.

\subsubsection*{4) Our understanding of human-AI interaction should include psychological and contextual factors.}

How people believe, trust, and rely on AI predictions for their decision-making should be seen in a larger context \cite{lee_trust_2004}. Contextual factors, including individual differences, task domain, information provided, media discourse, organizational and social environment, and more are all factors that could influence user perceptions and behavior in AI-assisted decision-making \cite{kordzadeh2022algorithmic, steyvers_three_2023}. 

Our findings emphasize that beyond system performance, user-related factors are critical in evaluating human-AI interaction. Psychological factors, including prior attitudes toward AI, paranormal beliefs, personality traits, and interest in the topic, significantly influence how people perceive AI predictions. These findings must be considered within the broader societal narratives, mental models, and cognitive biases that shape human-AI interactions.

Future research should prioritize interdisciplinary collaboration to delve deeper into the psychological and cultural dimensions of human-AI interactions. Psychologists and cognitive scientists can identify cognitive biases that shape user behavior, while sociologists and anthropologists can explore the influence of societal narratives on public perceptions of AI. Policymakers should actively support such research to better understand the broader societal implications of human-AI interactions. Additionally, AI system designers and developers must integrate these insights to create systems that align with realistic user expectations, fostering trust and effective collaboration between humans and AI.

\section{Limitations and Future Work}

First, our experiment focused on predictions about personal behavior to facilitate comparisons with astrology and personality-based predictions. This specific design may limit generalizability. Future research should explore perceptions of AI predictions in diverse contexts and domains, such as health, education, and workplace decision-making, as well as across different interaction settings (e.g., implicit versus explicit predictions). Examining psychological and cognitive factors in these areas could yield richer insights into belief formation and trust dynamics.

Second, our findings highlight the need for research into how cognitive style interacts with context-specific factors, such as perceptions of reliability and usefulness, in shaping beliefs about predictions. The significant interaction observed, which nearly reversed the nonsignificant main effect of validity, suggests that analytic thinkers may evaluate predictions flexibly rather than uniformly. Future studies could explore whether analytic cognitive styles involve distinct evaluation strategies for different prediction sources or whether contextual features, like framing or perceived expertise, influence these judgments. Understanding these mechanisms could inform interventions aimed at enhancing critical thinking and improving discernment between credible and pseudoscientific claims.

Third, our study primarily investigated user-related factors, leaving external influences such as social and organizational environments relatively unexplored. Future research could investigate how these broader contexts shape user perceptions, engagement, and trust in AI predictions, particularly in collaborative or high-stakes decision-making scenarios.

Finally, like most survey-based studies, ours may be affected by biases inherent to self-reported data, including social desirability bias \cite{nederhof_methods_1985} and common method bias \cite{podsakoff2003common}. To address these limitations, future studies could incorporate alternative methodologies, such as behavioral experiments or longitudinal designs, to validate and extend our findings.

\section{Conclusion}

Through a study with 238 participants, this research empirically investigates the phenomenon of “rational superstition” in AI. It provides a side-by-side comparison of people's perceptions of fictitious predictions based on AI, astrology, and personality. The results suggests that even without any validation of system
performance, people perceived fictitious AI predictions as highly believable, and their perceptions were positively associated with their perception of astrology- and personality-based predictions. Our findings showed that people's belief in AI predictions was positively associated with paranormal beliefs, positive attitude towards AI, interest in the topic of prediction, and negatively associated with conscientiousness and age. We also explored interactions with other prediction sources and subscales of believability, such as perceived validity, reliability, usefulness, and personalization. In our discussion, we highlighted the role of mental models and cognitive biases in shaping people's perceptions of AI predictions, and the importance of including psychological and contextual factors in studying human-AI interaction.

\section{Methods}
\label{sec:methods}

\subsection{Participants}
Data were initially collected from a survey administered to 300 participants recruited through Prolific. After excluding 62 participants due to incomplete responses or failed attention checks, the final analysis was conducted with data from the remaining 238 participants. These individuals were adults aged 18 and above from diverse socioeconomic backgrounds (see Table \ref{table:demographics} for an overview of participants' demographic information). 

\begin{table}[]
\centering
\caption{Summary of participants’ demographic information}
\vspace{10pt}
\label{table:demographics}
\begin{tabular}{@{}llc@{}}
\toprule
\multirow{2}{*}{Number per condition} & Positive     & 119   \\  
                                      & Negative     & 119   \\ \midrule
\multirow{2}{*}{Age}                  & Mean         & 41.04 \\  
                                      & SD           & 12.38 \\ \midrule
\multirow{3}{*}{Gender}               & Female       & 53\%  \\  
                                      & Male         & 43\%  \\  
                                      & Other        & 3\%   \\ \midrule
\multirow{5}{*}{Education}            & High school  & 12\%  \\  
                                      & Some college & 21\%  \\  
                                      & Bachelor     & 43\%  \\  
                                      & Master       & 14\%  \\  
                                      & Other        & 9\%   \\ \bottomrule
\end{tabular}
\end{table}

\subsection{Experiment Protocol}

\begin{figure}[h]
\includegraphics[width=15cm]{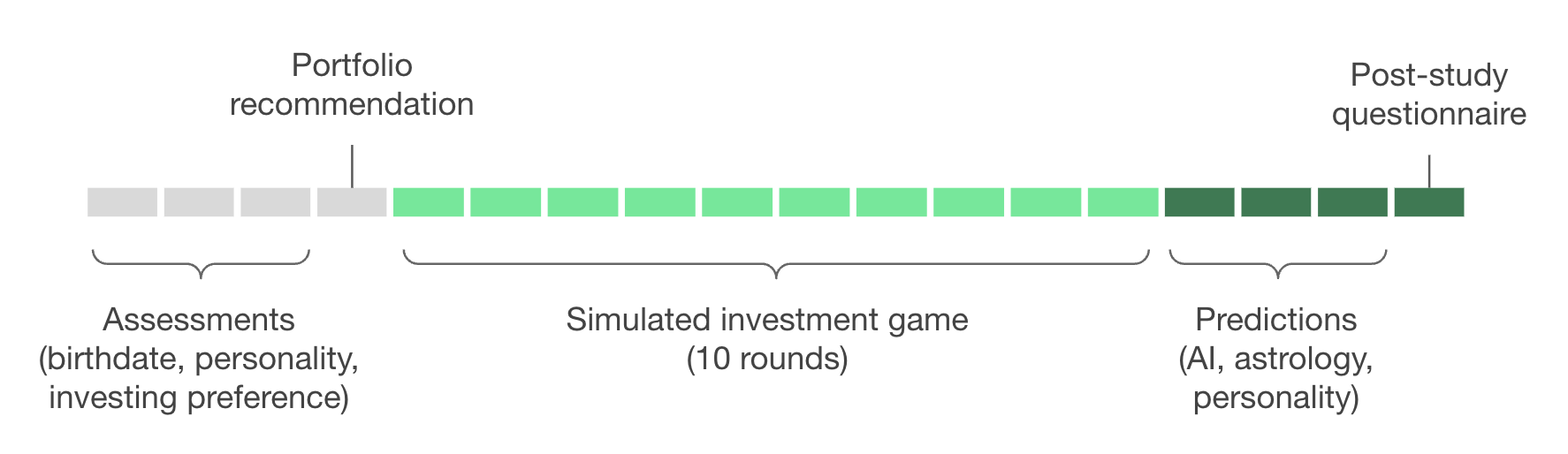}
    \centering
    \caption{Diagram of the study procedure}
\label{fig:procedure_diagram}
\end{figure}

\subsubsection{Assessments and simulated investment game}

Participants were asked to complete a short zodiac-related questionnaire (date, hour, and location of birth), a short personality test (adaptation of Myers-Briggs Type Indicator (MBTI) \cite{myers_myers-briggs_1962}, and a simulated investment game. In the instructions, participants were told that they will receive personalized predictions about their future investment behavior generated from three distinct sources (astrology, personality, and AI) based on their game interactions and the information they provided. What they were \emph{not} told in the beginning was that these predictions were generic, pre-determined statements that were not based on their responses, which was revealed to them after completion.

The simulated investment game was designed to elicit interactions comparable to that of modern robo-advising platforms \cite{dacunto_promises_2019}, but in a simplified way. Participants were provided with virtual currency of \$10,000 to invest across three investment categories (high risk/return, medium risk/return, low risk/return) over ten rounds (representing years). At each round, they could allocate up to 100\% of their assets across the three categories, with the goal of maximizing their total portfolio value. 

A few design choices were made to increase the sense of realism for the investing game and to increase engagement: 
\begin{itemize}
    \item \textbf{Recommended portfolio allocation:} Participants filled out a short questionnaire about their risk preferences for investing before starting the game and received a "customized" portfolio recommendation that they could view throughout the game. In reality, all participants received the same portfolio recommendation. 
    \item \textbf{Realistic market scenario:} The game was designed so that participants could experience the hypothetical ups and downs of the market, through a pre-determined series of alternating "bull market" (more gains) and "bear market" (more losses) scenarios that influenced the probability of the possible investment outcomes at each round.
    \item \textbf{Market forecasts:} To further enhance the emotional engagement, each round started off with either a positive or negative market forecast that were aligned with the behind-the-scenes market scenario. A detailed market scenario that was used in the game and examples of the market forecast can be found in the appendix [ref]. 
    \item \textbf{Financial incentives:} To motivate engagement, participants were told their investment performance would increase their total compensation for the study, inspired by previous studies that used real-life incentives to increase engagement \cite{wang_emotion_2014, shiv_investment_2005}.
\end{itemize}
 
\begin{figure}[h]
    \includegraphics[width=13cm]{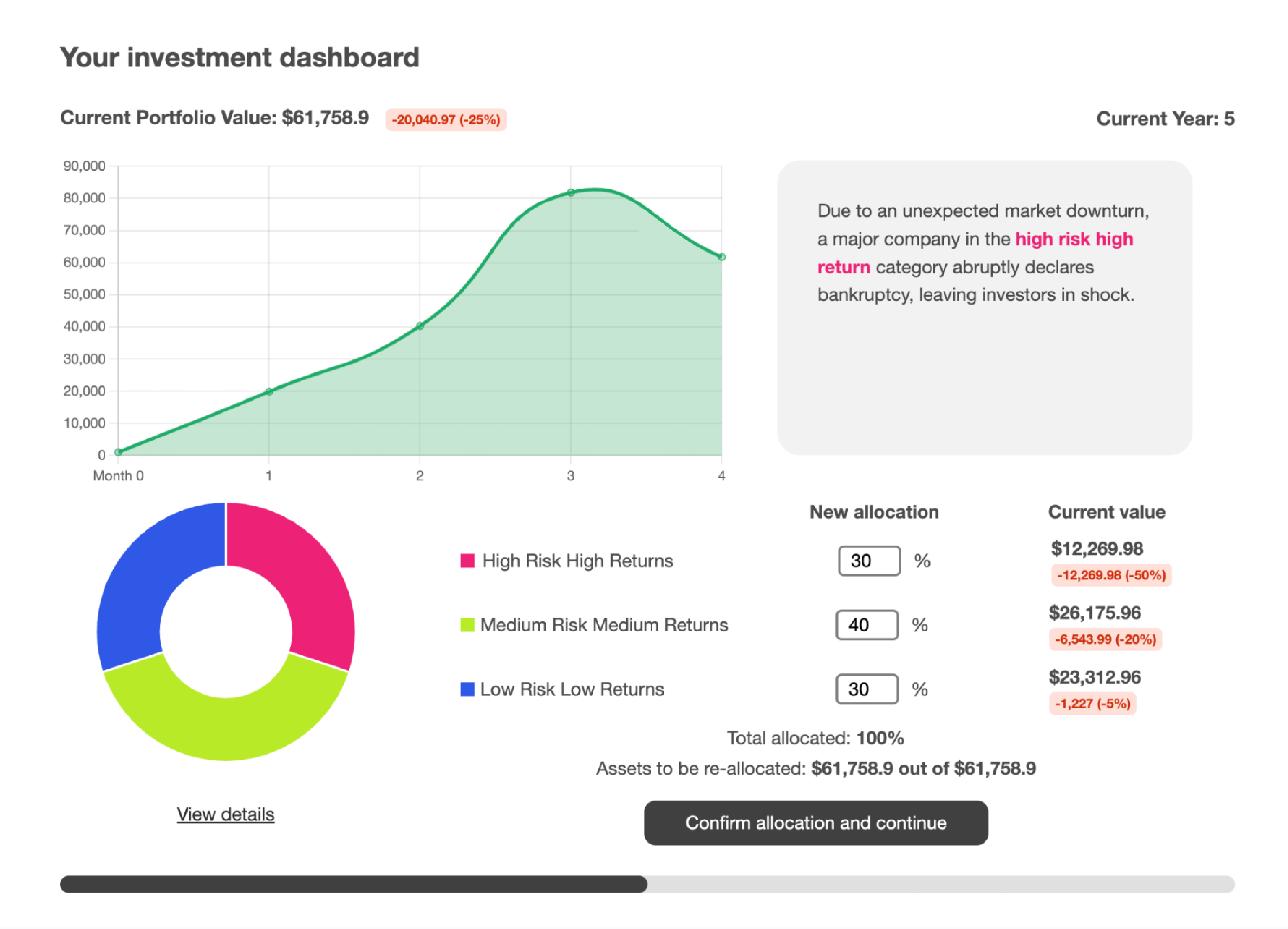}
    \centering
    \caption{Dashboard of the simulated investment game}
    \label{fig:game_dashboard}
\end{figure}

\subsubsection{Predictions and believability evaluation}

After the game, participants were presented with pre-determined, fictitious predictions from three different sources (AI, astrology, personality) in randomized order. Participants were randomly assigned to either the “Positive” prediction group (rational investor, higher returns) or “Negative” prediction group (impulsive investor, lower returns). Examples of predictions received are shown in Table \ref{table:predictions}. The initial predictions were written by the authors, with variations generated using ChatGPT-4. These variations were randomly presented to enhance variability and simulate AI-generated outputs, extending previous research on placebo effects of AI \cite{kloft2024ai, kosch_placebo_2022, villa_placebo_2023, villa_evaluating_2024}. For each prediction, participants were asked to evaluate its perceived validity, personalization, reliability, and usefulness. The evaluation questionnaire and rationale are provided in Section \ref{sec: believability_subscales}. 

\begin{table}[]
\centering
\caption{Example of predictions received by Positive and Negative prediction groups}
\vspace{10pt}
\label{table:predictions}
\begin{tabular}{@{}p{2cm}p{6.5cm}p{6.5cm}@{}}
\toprule
\textbf{Source} &
  \textbf{Positive prediction group} &
  \textbf{Negative prediction group} \\ \midrule
\textbf{Astrology} &
  As the Moon's celestial path brings it into close proximity with the stoic Saturn, this cosmic alignment sheds light on the inherent human struggle between emotion and reason when making financial decisions. Based on your astrological sign, you are more likely to make rational investment decisions based on reason rather than emotion. You have a good grasp of your emotions especially when it comes to decisions about money. Therefore, we predict that your portfolio will have higher-than-average performance in the long run. &
  As the Moon's celestial path brings it into close proximity with the stoic Saturn, this cosmic alignment sheds light on the inherent human struggle between emotion and reason when making financial decisions. Based on your astrological sign, your investment decisions are often impulsive, swayed by emotions rather than thoughtful analysis. Your emotions play a significant role in your investment decisions. As a result, it's expected that your portfolio's performance will be under the average in the long run. \\ \midrule
\textbf{Personality} &
  Whether you're an analytical thinker or led by intuition, your Myers-Briggs profile offers valuable insights into your behavior when it comes to making investment decisions. Based on your personality analysis, you are more likely to make rational investment decisions based on reason rather than emotion. You have a good grasp of your emotions especially when it comes to decisions about money. Therefore, we predict that your portfolio will have higher-than-average performance in the long run. &
  Whether you're an analytical thinker or led by intuition, your Myers-Briggs profile offers valuable insights into your behavior when it comes to making investment decisions. Based on your personality analysis, your investment decisions are often impulsive, swayed by emotions rather than thoughtful analysis. Your emotions play a significant role in your investment decisions. As a result, it's expected that your portfolio's performance will be under the average in the long run. \\ \midrule
\textbf{AI} &
  Within the digital echo of your actions, AI sifts through the noise to forecast your next move. It transforms your clicks, likes, and digital whispers into a map of future investment choices. Based on our AI model analysis, you are more likely to make rational investment decisions based on reason rather than emotion. You have a good grasp of your emotions especially when it comes to decisions about money. Therefore, we predict that your portfolio will have higher-than-average performance in the long run. &
  Within the digital echo of your actions, AI sifts through the noise to forecast your next move. It transforms your clicks, likes, and digital whispers into a map of future investment choices. Based on our AI model analysis, your investment decisions are often impulsive, swayed by emotions rather than thoughtful analysis. Your emotions play a significant role in your investment decisions. As a result, it's expected that your portfolio's performance will be under the average in the long run. \\ \bottomrule
\end{tabular}
\end{table}

\subsubsection{Post-study questionnaire}

Following the experiment, participants were asked to respond to a series of questionnaires to measure key user characteristics (cognitive style, paranormal beliefs, gullibility, trust in AI/attitudes towards AI, personality, etc.) and demographic information (age, gender, education level). A summary of the scales used in the questionnaires is provided in Table \ref{table:measures}. Detailed questionnaires can be found in Section \ref{sec:scales}.

\begin{table}[]
\centering
\caption{Summary of the measures for the user characteristics}
\vspace{10pt}
\label{table:measures}
\begin{tabular}{@{}p{4cm}p{11.5cm}@{}}
\toprule
\textbf{Variable name} & \textbf{Measure} \\ \midrule
Cognitive Style & A composite measure of cognitive style that combines cognitive reflection test (CRT-2) \cite{thomson_investigating_2016} and need for cognition (NCS-6) \cite{lins_de_holanda_coelho_very_2020}. Higher score indicates more analytic cognitive style. \\ \midrule
Paranormal Beliefs & A measure of participants’ belief in paranormal phenomena. A shortened version of R-PBS \cite{tobacyk_revised_2004} was used. Higher score indicates stronger belief in paranormal phenomena. \\ \midrule
AI Attitude Score & A score indicating participants’ attitudes towards AI predictions. AIAS-4 \cite{grassini_development_2023}, a 4-item scale was used. Higher score indicates more positive attitudes. \\ \midrule
Gullibility & A self-reported measure of participants' tendency to be gullible, developed by Teunisse et al. \cite{teunisse_i_2020}. Higher score indicates greater gullibility. \\ \midrule
Big Five Personality Traits & Includes extraversion, openness, agreeableness, conscientiousness, and emotional stability. Used 10-item Big Five Personality Inventory (TIPI) \cite{gosling_very_2003}. \\ \midrule
Interest in behavior & A measure of participants’ interest in the topic of prediction (personal investment behavior). "How would you rate your level of interest in understanding your future investment behavior?" (5-point Likert scale, 1=Not interested, 5=Extremely interested) \\ \midrule
Familiarity & A measure of participants' familiarity with the prediction sources (AI, Astrology, Personality). "How would you rate your level of familiarity with the following: 1) Astrology and horoscopes, 2) Personality psychology, 3) AI prediction systems" (5-point Likert scale, 1=Not at all, 5=Extremely familiar) \\ \bottomrule
\end{tabular}
\end{table}

\subsection{Measures}
\label{sec:scales}

This section includes the detailed questionnaires that were used in the study.

\subsubsection{Perceived validity, reliability, usefulness, and personalization}
\label{sec: believability_subscales}

To measure the believability of predictions, we employed four subscales: perceived validity, reliability, usefulness, and personalization. These subscales were informed by theoretical and empirical research on persuasion, technology acceptance, and user attitudes. Together, they capture distinct but interrelated dimensions of how users evaluate predictions from various sources.

\textbf{Perceived validity} reflects the degree to which users find predictions logically sound and factually accurate. This construct is rooted in the Elaboration Likelihood Model (ELM), which proposes that when individuals are motivated and able to process information, they engage in the central route of persuasion, where the quality and validity of arguments are critical determinants of attitude change \cite{petty_elaboration_1986}.

\textbf{Perceived reliability} measures the extent to which users view the source of the predictions as trustworthy and dependable. The concept of source reliability aligns with theories of source credibility, which identify trustworthiness and expertise as core elements influencing users’ acceptance of information \cite{hovland_communication_1953}. Within the ELM framework, the reliability of the source operates as a peripheral cue when individuals are not motivated to critically evaluate the content \cite{petty_elaboration_1986}.

\textbf{Perceived usefulness} reflects the degree to which users believe that the predictions can enhance their decision-making or outcomes. This dimension is derived from the Technology Acceptance Model (TAM), where perceived usefulness significantly influences attitudes and intentions toward technology adoption \cite{davis_perceived_1989}. In the context of predictions, if individuals find the information relevant and applicable, they are more likely to deem it believable because they address users’ goals and needs.

\textbf{Perceived personalization} captures the extent to which users perceive the predictions as tailored to their unique context or characteristics. Personalized communication has been shown to enhance believability by increasing attention and positive attitudes toward the message \cite{maslowska_it_2016}. Personalization also increases persuasion by making users feel understood and valued, and creates a sense of relevance, which can reduce cognitive load and enhance user engagement \cite{zhang_proactive_2019, kaptein_personalizing_2015}.

We asked participants to rate eight statements on 7-point Likert scales (1=Strongly disagree, 7=Strongly agree), which include two statements per subscale (perceived validity, personalization, reliability, and usefulness).

\begin{itemize}
    \item I find the prediction convincing. (validity\_1)
    \item I can identify with the prediction. (personalization\_1)
    \item The source of the prediction is reliable. (reliability\_1)
    \item I find the prediction helpful. (usefulness\_1)
    \item The prediction is accurate. (validity\_2)
    \item The prediction describes me very well.  (personalization\_2)
    \item I trust the source of the prediction. (reliability\_2)
    \item The prediction is useful for making future decisions. (usefulness\_2)
\end{itemize}

\subsubsection{Cognitive Reflection (CRT-2) and Need for Cognition (NCS-6)}

Cognitive reflection, or the tendency to suppress an intuitive but incorrect response in favor of a more reflective and correct one, was chosen as a moderating factor for belief in predictions to explore how it influences the acceptance and trust in predictions about personal behavior. This approach aims to determine if higher cognitive abilities correlate with more critical evaluation and discernment of AI predictions, as well as predictions based on astrology and personality.

The Cognitive Reflection Test (CRT) was designed to predict performance in normative decision-making \cite{frederick_cognitive_2005}. In our study, we used CRT-2 \cite{thomson_investigating_2016}, an alternate form of the original 6-item CRT, in order to address issues of familiarity and learning effects that could potentially skew results. The 4-item CRT-2 maintains the same objective of measuring cognitive reflection but includes different questions to ensure a fresh and unbiased assessment, with less focus on numeric abilities \cite{thomson_investigating_2016}. 

The Need for Cognition (NFC) test is a self-report measure of an individual's tendency to engage in and enjoy effortful thinking \cite{cacioppo_need_1982}. We use the 6-item NFC assessment (NCS-6) developed by Lins de Holanda Coelho et al. \cite{lins_de_holanda_coelho_very_2020}.

While the CRT measures objective performance, the NFC measures subjective preference, which have been found to be correlated \cite{pennycook_is_2016}. To account for the conceptual similarity of these two measures, they were combined to create a composite measure under the assumption that they have equal contribution to the final score. To do so, we summed the z-scores of the results from the two tests to account for the differences in range and variability \cite{field_discovering_2012}.

\textbf{Cognitive Reflection Test (CRT-2) \cite{thomson_investigating_2016}}:

\begin{itemize}
    \item If you’re running a race and you pass the person in second place, what place are you in? 
    \item A farmer had 15 sheep and all but 8 died. How many are left?
    \item Emily’s father has three daughters. The first two are named April and May. What is the third daughter’s name?
    \item How many cubic feet of dirt are there in a hole that is 3’ deep x 3’ wide x 3’ long?
\end{itemize}

\textbf{Need for Cognition (NCS-6) \cite{lins_de_holanda_coelho_very_2020}}:

\begin{itemize}
    \item I would prefer complex to simple problems. 
    \item I like to have the responsibility of handling a situation that requires a lot of thinking. 
    \item Thinking is not my idea of fun. (R) 
    \item I would rather do something that requires little thought than something that is sure to challenge my thinking abilities. (R) 
    \item I really enjoy a task that involves coming up with new solutions to problems. 
    \item I would prefer a task that is intellectual, difficult, and important to one that is somewhat important but does not require much thought.

\end{itemize}

\subsubsection{Paranormal beliefs (R-PBS)}

Paranormal beliefs refer to the conviction in phenomena beyond scientific explanation, such as supernatural events, astrology, and mystical experiences. Paranormal beliefs were chosen as a moderating factor for belief in AI, astrology, and personality predictions to investigate how these unconventional beliefs influence the acceptance of various predictive statements. 

To quantify the degree of paranormal belief, we adapted the Revised Paranormal Belief Scale (R-PBS) \cite{tobacyk_revised_2004} (7-point Likert scale, 1=Strongly disagree, 7=Strongly agree) to focus on the more relevant factors for our study, including traditional religious beliefs (4), spiritualism (4), precognition (4), superstition (3).

\begin{itemize}
    \item The soul continues to exist though the body may die. [religious beliefs]
    \item Black cats can bring bad luck. [superstition]
    \item Your mind or soul can leave your body and travel (astral projection). [spiritualism]
    \item Astrology is a way to accurately predict the future. [precognition]
    \item There is a devil. [religious beliefs]
    \item If you break a mirror, you will have bad luck. [superstition]
    \item During altered states, such as sleep or trances, the spirit can leave the body. [spiritualism]
    \item The horoscope accurately tells a person’s future. [precognition]
    \item I believe in God. [religious beliefs]
    \item The number “13” is unlucky. [superstition]
    \item Reincarnation does occur. [spiritualism]
    \item Some psychics can accurately predict the future. [precognition]
    \item There is a heaven and a hell. [religious beliefs]
    \item It is possible to communicate with the dead. [spiritualism]
    \item Some people have an unexplained ability to predict the future. [precognition]

\end{itemize}

\subsubsection{AI Attitude/Trust in AI (AIAS-4)}

A 4-item, 1-factor AI attitude scale (AIAS-4) \cite{grassini_development_2023} (10-point Likert scale, 1=Not at all, 10=Completely agree) was used to measure participants' attitude towards AI. 

\begin{itemize}
    \item I believe that AI will improve my life.
    \item I believe that AI will improve my work.
    \item I think I will use AI technology in the future.
    \item I think AI technology is positive for humanity.
\end{itemize}

\subsubsection{Gullibility}

A shortened version of the originally 12-item, 2-factor scale by Teunisse et al. (2020) \cite{teunisse_i_2020} (7-point Likert scale, 1=Strongly disagree, 7=Strongly agree) was used to measure self-reported gullibility. 

\begin{itemize}
    \item (G4) I’m not that good at reading the signs that someone is trying to manipulate me. [insensitivity]
    \item (G5) I’m pretty poor at working out if someone is tricking me. [insensitivity]
    \item (G6) It usually takes me a while to “catch on” when someone is deceiving me. [insensitivity]
    \item (G1) I guess I am more gullible than the average person. [persuadability]
    \item (G9) My friends think I’m easily fooled. [persuadability]
    \item (G12) Overall, I’m pretty easily manipulated. [persuadability]

\end{itemize}

\subsubsection{Big Five Personality}

Based on the Five Factor personality model, the 10-item Big Five Personality Inventory (TIPI) \cite{gosling_very_2003} (7-point Likert scale, 1=Strongly disagree, 7=Strongly agree) was used to measure participants' personality.


I see myself as:

\begin{itemize}
    \item Extraverted, enthusiastic.
    \item Critical, quarrelsome.
    \item Dependable, self-disciplined.
    \item Anxious, easily upset.
    \item Open to new experiences, complex.
    \item Reserved, quiet.
    \item Sympathetic, warm.
    \item Disorganized, careless.
    \item Calm, emotionally stable.
    \item Conventional, uncreative.
\end{itemize}

\subsubsection{Familiarity/Level of expertise}

How would you rate your level of familiarity with the following (5-point Likert scale, 1=Not at all, 5=Extremely familiar)
\begin{itemize}
    \item Astrology and horoscopes
    \item Personality psychology
    \item AI prediction systems
\end{itemize}

\subsubsection{Interest in topic of prediction}

How would you rate your level of interest in understanding your future investment behavior? (5-point Likert scale, 1=Not interested, 5=Extremely interested)

\subsection{Approvals} 
This research was reviewed and approved by the MIT Committee on the Use of Humans as Experimental Subjects, protocol number E-5768. The study was pre-registered on AsPredicted.org (see \url{https://aspredicted.org/QDF_8KX}).

\subsection{Analysis}
\label{sec:analysis}

\subsubsection{Variables}
\label{subsec:variables}

The dependent variable (Y) was the subscale scores for perceived validity, personalization, reliability, and usefulness (7-point Likert scale, 1=Strongly disagree, 7=Strongly agree). 

The analysis included a range of predictor variables (X) to examine their effects on the subscale scores. The predictor and control variables were:

\begin{itemize}
    \item \textit{prophecy\_source}: A categorical variable indicating the source of the prophecy (Astrology, Personality).
    \item \textit{prophecy\_group}: A categorical variable indicating the valence of prophecy (Positive, Negative).
    \item \textit{composite\_score}: A composite measure of cognitive style.
    \item \textit{paranormal\_score}: A measure of participants' belief in paranormal phenomena. 
    \item \textit{aias\_score}:  A score indicating participants' attitudes/trust towards AI predictions.
    \item \textit{gullibility\_score}: A self-reported measure of participants' tendency to be gullible.
    \item Big Five Personality Traits: A measure including extraversion, openness, agreeableness, conscientiousness, and emotional stability.
    \item \textit{interest\_behavior}: A measure of participants' interest in the topic of prediction (personal investment behavior).
    \item \textit{familiarity}: A measure of participants' familiarity with the prediction sources (AI, Astrology, Personality).
    \item \textit{age}
    \item \textit{gender}: A categorical variable (Female, Male, Other)
    \item \textit{education}: A categorical variable (Bachelor, Master, Doctorate, Professional degree, Associate, Some college, High school, Less than high school)
\end{itemize}

\subsubsection{Multiple linear regression}

To test the hypothesis that individuals who are more likely to believe in astrology and personality-based predictions are also more likely to believe in AI predictions (H1), a multiple linear regression model was first applied to the data in wide format. The dependent variable was \textit{ai\_overall\_score}, and the main predictors were \textit{zodiac\_overall\_score} and \textit{personality\_overall\_score}, with the rest of the variables mentioned in Section \ref{subsec:variables} included as control variables. The model was fitted using the lm function in R.

\begin{equation}
\text{ai\_overall\_score} \sim \text{zodiac\_overall\_score} + \text{personality\_overall\_score} + \text{control variables}
\end{equation}

The following model diagnostics were performed to ensure the validity of the model:

\begin{itemize}
    \item \textbf{Linearity}: Residual plots for the linear regression model indicated no clear patterns, suggesting linearity.
    \item \textbf{Homoscedasticity}: The residual vs. fitted values plot for the linear regression model showed some heteroscedasticity, which was addressed in the mixed-effects model.
    \item \textbf{Normality of Residuals}: The Q-Q plot indicated that the residuals were approximately normally distributed.
    \item \textbf{Multicollinearity}: Variance Inflation Factors (VIFs) were calculated for the linear regression model, indicating that multicollinearity was not a concern. 
    \item \textbf{Independence of Residuals}: The Durbin-Watson test for the linear regression model showed no significant autocorrelation. 
\end{itemize}

\subsubsection{Mixed effects model}
\label{subsubsec:mixed_effects_model}

Mixed-effects models are well-suited for hierarchical data, incorporating both fixed and random effects to test multiple hypotheses simultaneously. They use partial pooling, which shifts group estimates toward the overall mean, effectively reducing Type I errors without the severe power loss of traditional multiple comparisons corrections. This approach accounts for data dependencies and correlations, minimizing the need for explicit adjustments for multiple comparisons \cite{gelman2012why}.

To examine the main effects and moderating effects of psychological and cognitive factors (H2, H3), the data was transformed to a long format, creating a three-level hierarchical structure. Each subject engaged in three types of conditions (\textit{prophecy\_source}) and evaluated across four subscales (\textit{subscale}). The outcome variable was the \textit{subscale\_score}.

The mixed effects model was specified as follows. The fixed effects were defined as:

\begin{equation}
\begin{aligned}
\text{subscale\_score} \sim & \, \text{subscale} * (\text{prophecy\_source} * \text{prophecy\_group} \\
& + \text{prophecy\_source} * \text{composite\_score} + \text{prophecy\_source} * \text{paranormal\_score} \\
& + \text{prophecy\_source} * \text{aias\_score} + \text{prophecy\_source} * \text{gullibility\_score} \\
& + \text{big5\_extraversion} + \text{big5\_openness} + \text{big5\_agreeableness} + \text{big5\_conscientiousness} \\
& + \text{big5\_emotional\_stability} + \text{interest\_behavior} + \text{familiarity} \\
& + \text{prophecy\_source} * \text{Age} + \text{education} + \text{prophecy\_source} * \text{gender})
\end{aligned}
\end{equation}

For the random effects, random intercepts were included to account for variability in baseline levels between subjects (\textit{qualtrics\_code}), and random slopes were included to account for variability in how subjects respond to different conditions (\textit{prophecy\_source}):

\begin{equation}
\text{random  } = \quad \sim \text{prophecy\_source} \mid \text{qualtrics\_code}
\end{equation}

Additionally, the model incorporated a correlation structure and variance weights to further account for the hierarchical structure and heteroscedasticity in the data:

\begin{equation}
\begin{aligned}
\text{correlation} &= \text{corSymm(form = ~1 | qualtrics\_code/prophecy\_source)}, \\
\text{weights} &= \text{varIdent(form = ~1 | subscale * prophecy\_source * prophecy\_group)}
\end{aligned}
\end{equation}

We centered the continuous predictor variables to reduce multicollinearity and improve the stability and interpretability of the coefficient estimates. Centering involved subtracting the mean of the variable from each individual value. This transforms the variable to have a mean of zero while preserving its variance and distribution. After centering the predictors, we observed that four previously non-significant main and interaction terms became significant, including the main effect of Personalization subscale, the main effect Astrology prediction source, and interaction terms between subscale (Personalization, Reliability) and Astrology prophecy source. This change is attributed to the reduction in multicollinearity, resulting in more precise coefficient estimates.

The model was implemented using the lme function from the nlme package in R. The analysis controlled for missing values through listwise deletion. Optimization settings were adjusted to ensure convergence, with maximum iterations and evaluations set to 1000.

Several model diagnostics were performed to ensure the validity of the model:

\begin{itemize}
    \item \textbf{Linearity}: A residuals vs fitted values plot using standardized residuals indicated no clear patterns, suggesting linearity.
    \item \textbf{Homoscedasticity}: A residuals vs fitted values plot showed a random spread around the horizontal line at zero, confirming that heteroscedasticity was successfully addressed using the variance function (varIdent). One limitation to this diagnostics was the less continuous nature of the outcome variable (\textit{subscale\_score}), which was multilevel with 13 levels (value of 1 to 7 with an increment of 0.5), which caused the residual plot to exhibit a slight diagonal pattern.
    \item \textbf{Normality of Residuals}: Normality of residuals was assessed using a Q-Q plot, using standardized residuals to account for both fixed and random effects. While there were slight deviations at the extremes, they were not severe and thus considered as normal deviations from real-world data. 
    \item \textbf{Multicollinearity}: 
    To address potential structural multicollinearity, we centered the continuous predictor variables. This process reduced the Variance Inflation Factor (VIF) values for all fixed effects to below 5, indicating that multicollinearity was reduced to an acceptable level. The model diagnostics, including standard errors and confidence intervals, confirmed the stability and reliability of the coefficient estimates.
    
    \item \textbf{Independence of Residuals}: 
    We applied the Durbin-Watson test at the subscale level to check for autocorrelation, finding significant positive autocorrelation in each group. Initially, we used an autoregressive correlation structure (corAR1), which resolved the issue for the "Validity" subscale and slightly reduced autocorrelation in others. However, model fit statistics suggested that an unstructured symmetric correlation structure (corSymm) was a better fit, though it did not fully address the positive autocorrelation. Comparison of the p-values, standard errors, and estimates between models with corAR1 and corSymm structures revealed no notable differences. We also attempted to model the order effect of the prophecies and subscales to mitigate this autocorrelation, but this approach did not lead to notable improvements. This unresolved autocorrelation may limit the validity of our results, and future studies should explore alternative approaches or additional data collection to address this issue.
    
    \item \textbf{Normality of Random Effects}: A Q-Q plot of random effects was used to assess normality. The points closely followed the reference line, indicating overall normality, with slight deviations at the tails suggesting minor departures. However, these deviations were not significant enough to challenge the model assumptions.

    \item \textbf{Random Effects Evaluation:} The variance components and Intraclass Correlation Coefficient (ICC) were analyzed to evaluate the random effects structure. The ICC values (Adjusted ICC: 0.939, Unadjusted ICC: 0.598) indicated that a substantial portion of the variance was due to random effects. To further assess their contribution to overall model variability, we examined the variance components in the mixed effects model, which included random intercepts and slopes for \textit{prophecy\_source} at the \textit{qualtrics\_code} level. The estimated variance components are presented in Table \ref{table:random_effects_correlations}. 

\end{itemize}

\begin{table}[h]
\centering
\caption{Random Effects Variance, Standard Deviation, and Correlations}
\vspace{10pt}
\label{table:random_effects_correlations}
\begin{tabular}{@{}p{4.2cm}>{\centering\arraybackslash}p{1.6cm}>{\centering\arraybackslash}p{1.6cm}>{\centering\arraybackslash}p{1.6cm}>{\centering\arraybackslash}p{1.6cm}@{}}
\toprule
\textbf{Random Effect} & \textbf{Variance} & \textbf{SD} & \textbf{Corr w Intercept} & \textbf{Corr w Astrology} \\ \midrule
Intercept                   & 1.560 & 1.249 & -      & -     \\ 
prophecy\_sourceAstrology   & 1.211 & 1.100 & -0.260 & -     \\ 
prophecy\_sourcePersonality & 1.002 & 1.001 & -0.221 & 0.293 \\ 
Residual                    & 0.121 & 0.348 & -      & -     \\ \bottomrule
\end{tabular}
\end{table}

\section{Data Availability}

The datasets generated during and analysed during the current study are available in the GitHub repository, \url{https://github.com/mitmedialab/ai-superstition}.

\section{Code Availability}

The source code for the simulated investment game and analysis is available in the GitHub repository, \url{https://github.com/mitmedialab/ai-superstition}.

\section*{Acknowledgments}
We thank Jinjie Liu, Data Science Specialist at the Institute for Quantitative Social Science at Harvard University, for sharing her valuable perspective on the analysis.

\section{Author Contributions}

E.L. and P.P. conceptualized and designed the study. E.L. implemented the experiment, conducted the study, and performed data analysis and interpretation, with contributions from P.P. E.L. drafted the manuscript and P.P. provided critical revisions. J.A. and P.M. provided feedback throughout the process and edited the manuscript. All authors approved the final version of the manuscript.

\bibliographystyle{unsrt}  
\bibliography{references}

\end{document}